\documentstyle[preprint,revtex]{aps}
\begin{document}
\draft
\baselineskip = 1.1\baselineskip
\begin{title}
{Chiral transitions in three--dimensional magnets \\
and higher order $\epsilon$--expansion} 
\end{title}
\author{S.~A.~Antonenko, A.~I.~Sokolov, K.~B.~Varnashev}
\begin{instit}
Department of Physical Electronics, Saint Petersburg Electrotechnical
University, \\
Professor Popov Street 5, St.Petersburg, 197376, Russia
\end{instit}
\begin{abstract}
The critical behaviour of helimagnets and stacked triangular 
antiferromagnets is analyzed in $(4 - \epsilon)$ dimensions within
three--loop approximation. Numerical estimates for marginal values
of the order parameter dimensionality $N$ obtained by resummation
of corresponding $\epsilon$--expansions rule out the possibility
of continuous chiral transitions in magnets with Heisenberg or
planar spins. 

\vspace{2cm}

stacked triangular antiferromagnets.
\end{abstract}

\newpage

Chiral phase transitions in helical magnets and stacked
triangular antiferromagnets with Heisenberg or $XY$--like
spins as well
as in some other systems attract much attention during the last
decade \cite{1,2,3,4,5,6,7,8,9}. Special interest to these
transitions demonstrated
both by theorists and experimentalists has been given rise by the
conjecture \cite{1,2,3} that they belong to new universality
class which
is characterized by critical exponents differing markedly from
those of 3D $O(n)$--symmetric model with relevant values of
$n$ (2, 4, and 6). This conjecture originates from the
renormalization--group
(RG) analysis of corresponding $(4 - \epsilon)$--dimensional
model performed
within the lowest (one- and two-loop, according to quantum field
theory language) orders in $\epsilon$. Some results given by
the $1 \over n$--expansion
and Monte Carlo simulations were also considered as favoring
the abovementioned idea \cite{2,3}.

In this Letter, an attempt is made to clear up whether the
conclusion about an existence of new universality class for
3D chiral systems with Heisenberg or planar spins survives
when higher-order terms in $\epsilon$--expansion are taken
into account. Below we calculate $\epsilon$--expansions for
quantities of interest up to the three--loop order. To obtain
numerical estimates relevant to real 3D magnets we apply
resummation procedures to these series before setting
$\epsilon = 1$. Such a machinery proved to give good results
for plenty of phase transition models. It is believed to
be powerful enough to yield reasonable predictions in our
case as well.

The Landau--Wilson Hamiltonian describing systems under
consideration may be written down in the form (see, e.g.
Ref.~\cite{9}):
\begin{equation}
H = {1 \over 2}
\int d^D x \Bigl[ m_0^2 \varphi_{\alpha} \varphi_{\alpha}^*
 + \nabla \varphi_{\alpha} \nabla \varphi_{\alpha}^*
+ {u_0 \over 2} \varphi_{\alpha} \varphi_{\alpha}^* \varphi_{\beta}
\varphi_{\beta}^*
+ {w_0 \over 2} \varphi_{\alpha} \varphi_{\alpha} \varphi_{\beta}^*
\varphi_{\beta}^* \Bigr] \ \ , \label{eq:1}
\end{equation}
where $\varphi_{\alpha}$ is a complex vector order parameter field,
$\alpha, \beta = 1, 2,$ \ldots, $N$,
a bare mass squared $m_0^2$ being proportional to the deviation
from the mean--field transition point. This model undergoes chiral
phase transitions if $w_0 > 0$ \cite{9}. In the opposite case, the
transitions into somewhat trivial (linearly polarized or unfrustrated)
ordered states take place.

In the critical region, where fluctuations are strong and the
system behavior is governed by the RG equations, the model
Eq.~(\ref{eq:1})
can demonstrate four different regimes of RG flow depending on $N$
\cite{3,9}. Correspondingly, three critical (marginal) values
of $N$ exist
separating these regimes from each other. If $N < N_{c1}$
the RG equations
possess three nontrivial fixed points (FP's) with the
$O(2N)$--symmetric
point being stable. When $N$ exceeds $N_{c1}$ the Heisenberg
FP loses its stability but the other,
anisotropic FP with coordinate $w < 0$ acquires
it. This point ``annihilates'' with another,
saddle anisotropic FP when $N$ approaches $N_{c2}$,
and there is only one nontrivial FP in the domain
$N_{c2} < N < N_{c3}$. It is $O(2N)$--symmetric and unstable.
At last, when $N$
increases further and crosses over the value $N_{c3}$
the creation of two new anisotropic FP's with $w > 0$
takes place. One of them is stable
and describes the chiral critical behavior. Hence, to answer
the question about the relevance of the chiral FP to the critical
thermodynamics of real helical magnets and stacked triangular
antiferromagnets, one has to estimate $N_{c3}$ and compare the number
obtained with physical values $N = 2$ and $N = 3$.

Marginal values of $N$ may be found analyzing RG $\beta$--functions.
We calculate these functions for the model Eq.~(\ref{eq:1})
within three--loop approximation in $(4 - \epsilon)$ dimensions 
using the minimal subtraction scheme (corresponding expansions are 
too lengthy and not presented here). The $\epsilon$--expansion
for $N_{c3}$ resulting from the $\beta$--functions obtained is 
as follows:
\begin{eqnarray}
N_{c3} = 12 + 4{\sqrt 6} - \biggl(12 + {14 {\sqrt 6} \over 3}
\biggr) \epsilon + \biggl[ {137 \over 150} + {91 {\sqrt 6} \over 300}
+ \biggl( {13 \over 5} + {47 {\sqrt 6} \over 60} \biggr) \zeta (3)
\biggr] \epsilon^2 \nonumber \\
= 21.80 - 23.43 \epsilon + 7.088 \epsilon^2 \ \ ,
\qquad \qquad \qquad \qquad \qquad \qquad \qquad \qquad \quad \ \
\label{eq:2}
\end{eqnarray}
where $\zeta (x)$ is the Riemann $\zeta$--function,
$\zeta (3) = 1.20206$. The constant
and linear terms in Eq.~(\ref{eq:2}) coincide with those presented by
H.~Kawamura \cite{3} while the second--order one is essentially new.

Such expansions are known to be asymptotic and
physical information may be extracted from them provided
some resummation method is applied. The Borel transformation
combined with proper procedure of analytical continuation of the
Borel transform usually plays a role of this method leading to
precise numerical estimates in cases of long enough original
series \cite{11,12}. To perform the analytical continuation the
Pade approximant
of $[ L/1 ]$ type may be used which is known to provide rather good
results for various Landau--Wilson models
(see, e.g. Refs.~\cite{9,13,14}).
The Pade--Borel summation of the expansion (\ref{eq:2}) gives:
\begin{equation}
N_{c3} = a - {2 b^2 \over c} + {4 b^3 \over {c^2 \epsilon}}
exp \biggl( {- 2 b \over {c \epsilon}} \biggr)
Ei \biggl( {2 b \over {c \epsilon}} \biggr) \ \ ,
\label{eq:3}
\end{equation}
where $a$, $b$, and $c$ are coefficients before $\epsilon^0$,
$\epsilon^1$, and $\epsilon^2$ in Eq.~(\ref{eq:2}),
respectively, $Ei(x)$ being the exponential integral.
Setting $\epsilon = 1$, we obtain from Eq.~(\ref{eq:3})
\begin{equation}
N_{c3} = 3.39 \ \ . \label{eq:4}
\end{equation}
Making use of the Pade approximant $[ 1/1 ]$ itself gives
$N_{c3} = 3.81$ while direct summation of the expansion
Eq.~(\ref{eq:2}), being rather crude procedure, results
in $N_{c3} = 5.46$.

All these numbers, although considerably scattered, are
nevertheless greater than 3. Hence, helical magnets and stacked
triangular antiferromagnets with Heisenberg and $XY$--like spins
are not seen to demonstrate new, chiral critical behavior. Instead, 
they should approach helically ordered state or frustrated 
antiferromagnetic phase only via first--order phase transitions.

On the other hand, the difference between the number Eq.~(\ref{eq:4})
and $N = 3$ is not so large. Moreover, numerical estimates for $N_{c3}$
were obtained from the theory having no small parameter in the limit
$\epsilon = 1$. How close to the precise value of $N_{c3}$ they may be?

To clear up this point let us compare Eq.~(\ref{eq:4}) and its
Pade counterpart with analogous
estimate given by the RG analysis in three dimensions. Three--loop
RG expansions for $\beta$--functions resummed by means of
generalized Pade--Borel technique result in $N_{c3} = 3.91$ \cite{9}.
Since the accuracy provided by this approximation was argued
to be rather high \cite{9} (about $1 \%$ for FP's coordinates
and critical exponents), resummed $\epsilon$--expansion,
within three--loop order, seems to yield plausible estimates
for the quantities of interest.

This conclusion is of prime importance. It is reasonable, therefore, 
to look for extra pros and cons it. One can find them calculating 
the rest of marginal values of $N$ from resummed three--loop
$\epsilon$--expansions and making a comparison of numbers obtained with
corresponding 3D RG estimates. The inequality $N_{c2} > 2$ proven 
earlier \cite{9} may be used as a criterion in course of such a study. 
The $\epsilon$--expansions for $N_{c1}$ and $N_{c2}$ are found to be:
\begin{equation}
N_{c1} = 2 - \epsilon + {5 \over 24} \bigl(6 \zeta (3) - 1 \bigr)
\epsilon^2 = 2 - \epsilon + 1.294 \epsilon^2 \ \ , \label{eq:5}
\end{equation}
\begin{eqnarray}
N_{c2} =  12 - 4{\sqrt 6} - \biggl(12 - {14 {\sqrt 6} \over 3}
\biggr) \epsilon + \biggl[ {137 \over 150} - {91 {\sqrt 6} \over 300}
+ \biggl( {13 \over 5} - {47 {\sqrt 6} \over 60} \biggr) \zeta (3)
\biggr] \epsilon^2  \nonumber \\
= 2.202 - 0.569 \epsilon + 0.989 \epsilon^2 \ \ .
\qquad \qquad \qquad \qquad \qquad \qquad \qquad \qquad \quad \ \
\label{eq:6}
\end{eqnarray}
Their Pade--Borel summation gives for $\epsilon = 1$
\begin{equation}
N_{c1}  = 1.50 \ \ , \quad N_{c2} = 1.96 \ \ . \label{eq:7}
\end{equation}

These values are close to those obtained in 3D: $N_{c1} = 1.45,
N_{c2} = 2.03$ \cite{9}, but $N_{c2}$ is seen to be
obviously underestimated by the $\epsilon$--expansion
since corresponding number is less than 2. The
difference $2 - N_{c2}^{(\epsilon)} = 0.04$, however,
is small and may be considered
as a lower bound for the error produced by this approximation.
The most likely estimate for this error is believed to be
close to $N_{c2}^{\rm (3D)} - N_{c2}^{(\epsilon)}$,
i.e. being of order of 0.1.

For $N_{c1}$ the $\epsilon$--expansion predicts the value
which is slightly greater than $N_{c1}^{\rm (3D)}$.
At the same time, these numbers differ from
each other by only $3 \%$ and lie so far from the nearest physical
value $N = 2$ that this difference is quite unimportant.

We see that resummed three--loop $\epsilon$--expansion gives good
enough numerical estimates for $N_{c1}$ and $N_{c2}$ providing,
however, the lower accuracy when used for evaluation
of highest critical dimensionality $N_{c3}$. It is not
surprising since the structure of series Eq.~(\ref{eq:2})
turns out to be rather unsuitable for yielding reliable
quantitative results. Indeed, to obtain precise numerical
estimates one has to deal with series which, at least, possess
coefficients decreasing with increasing their number.
Instead, the second term in Eq.~(\ref{eq:2}) exceeds,
for $\epsilon = 1$, the first one. It is clear that such
an expansion would not demonstrate good summability.
That is why we believe that the true value of $N_{c3}$ is
closer to 3D estimate 3.91 than to 3.39. On the other hand,
the latter estimate differs from the former by no more than
$15 \%$. Hence, actually the current status of the
$\epsilon$--expansion is not so bad provided three--loop
contributions are properly taken into account. It seems natural
that calculations of higher--order terms will result in further
improvement of numerical estimates.

It is worth noting that really three--loop terms added and the
summation have changed the situation drastically. The point is
that two--loop $\epsilon$--expansions for $N_{c1}, N_{c2}$,
and $N_{c3}$ directly extrapolated to $\epsilon = 1$ violate
the inequalities $N_{c1} < N_{c2} < N_{c3}$ which should hold
good according to the definition of $N_{ci}$; in this approximation
$N_{c3}^{(\epsilon)} < N_{c1}^{(\epsilon)} < N_{c2}^{(\epsilon)}$
when $\epsilon = 1$. This is an alarm bell signalizing that
with such short series in hand one can not safely penetrate into
the three--dimensional world. To the contrary, the numbers given by
the Pade and Pade--Borel resummed three--loop $\epsilon$--expansions
at $\epsilon = 1$ meet abovementioned inequalities.

Another point to be specially marked is as follows. Actually, we
have now enough information resulting from higher--order RG analysis 
in three and $4 - \epsilon$ dimensions to make the firm conclusion 
about the critical behavior of the $XY$--like systems. Indeed, 
since $N_{c2}$ has been proven to be greater than 2 \cite{9} and 
$N_{c3}$ should exceed $N_{c2}$, magnets with planar spins certainly 
can not undergo continuous chiral transitions. Only first-order 
phase transitions into chiral states are possible, in principle, 
in these systems.

We conclude with the comment concerning the $1 \over n$--expansion and
Monte Carlo simulations. Actually, their predictions do not contradict
to those just obtained. Indeed, as was shown in Ref.~\cite{14}, even
for simple, $O(n)$--symmetric model the $1 \over n$--expansion
begin to yield reasonable numerical estimates only when $n$ exceeds 20.
Since for systems studied $n = 2N = 4, 6$, this approach is obviously
inapplicable in our case. Monte Carlo simulations, in turn, give the
values of critical exponents which are close to tricritical ones,
especially for systems with $XY$--like spins \cite{15}.
That is why it is believed \cite{5,9,15} that not chiral critical
behavior but tricritical one or tricritical--to--critical crossover
are really seen in these computer experiments.

To summarize, we have found three--loop contributions to the
$\epsilon$--expansions of critical order--parameter
dimensionalities $N_{c1}$, $N_{c2}$, and $N_{c3}$ for the model
describing the chiral critical behavior. The Pade--Borel
summation of series obtained has yielded, at $\epsilon = 1$,
fair numerical estimates for $N_{c1}$ and $N_{c2}$ in 3D. For the
lower boundary of the domain where continuous chiral transitions
are possible the $\epsilon$--expansion resummed by Pade-Borel and
simple Pade methods has given $N_{c3} = 3.39$ and $N_{c3} = 3.81$
respectively. Being close enough to 3D RG estimate $N_{c3} = 3.91$,
these numbers are greater than physical values $N = 2$ and $N = 3$.
It may be considered as an evidence that in magnets with Heisenberg
or planar spins the chiral critical behavior with specific values of
critical exponents would not really occur and they can not belong 
to new, chiral class of universality. Since $N_{c3} > N_{c2}$ and
$N_{c2}$ should exceed 2, for the $XY$--like systems this conclusion 
sounds as firm.

This work was supported in part by the State Committee of
Russian Federation for Higher Education through Grant 
No. 94--7.17--351 and in part by the Foundation for Intellectual 
Collaboration (St.Petersburg) via Interdisciplinary Scientific 
and Technological Program of Russia ``Fullerenes and Atomic 
Clusters'', Project No. 24.

\end{document}